\documentclass[12pt,a4paper]{article}
\pdfoutput=1

\usepackage{geometry}
\usepackage{amsmath}
\usepackage{amsmath}
\usepackage{amssymb}
\usepackage[dvips]{graphicx}
\usepackage{cite}
\usepackage{booktabs}
\usepackage{pstricks}
\usepackage{bm}
\usepackage{pbox}
\usepackage{placeins}
\usepackage{graphicx}
\usepackage{caption}
\usepackage{subcaption}
\usepackage[T1]{fontenc}
\usepackage{footnote}
\usepackage{pdfpages}
\usepackage{hhline}
\usepackage{multirow}
\usepackage{multicol}
\usepackage{enumitem}
\usepackage{multirow}
\usepackage{amsmath}
\usepackage{relsize}
\usepackage{float}
 \usepackage{graphicx}
\usepackage[toc,page]{appendix}
\usepackage[english]{babel}
\allowdisplaybreaks

\usepackage{slashed}
\usepackage{cases}

\usepackage{empheq}

\definecolor{MyDarkBlue}{rgb}{0.1, 0.3, 0.8} %defining the color 'MyDarkBlue'
\definecolor{SBlue}{rgb}{0.2, 0.4, 0.4} %defining the color 'MyDarkBlue'
\definecolor{MyLightBlue}{rgb}{0.22,0.51,0.99}
\definecolor{MyGreen}{rgb}{0.0, 0.5, 0.3}
\definecolor{BrickRed}{rgb}{0.8, 0.25, 0.33}
\usepackage[colorlinks=true,linkcolor=blue,citecolor=MyDarkBlue,
urlcolor=BrickRed,bookmarksnumbered=true,bookmarksopen]{hyperref}
\hypersetup{colorlinks, citecolor=SBlue,linkcolor=MyGreen, urlcolor=MyDarkBlue}

\begin{document}
\vspace*{-0.2in}
\begin{flushright}
\end{flushright}
\renewcommand{\thefootnote}{\fnsymbol{footnote}}
\begin{center}
{\Large \bf
Gravitational Wave Imprints of Left-Right Symmetric \\ \vspace{0.1in} Model   with Minimal Higgs Sector
}
\end{center}
\renewcommand{\thefootnote}{\fnsymbol{footnote}}
\begin{center}
{
{}~\textbf{Luk\'{a}\v{s} Gr\'{a}f$^{1,2,3}$}\footnote{ E-mail: \textcolor{MyDarkBlue}{lukas.graf@berkeley.edu}},
{}~\textbf{Sudip Jana$^1$}\footnote{ E-mail: \textcolor{MyDarkBlue}{sudip.jana@mpi-hd.mpg.de}},
{}~\textbf{Ajay Kaladharan $^4$}\footnote{ E-mail: \textcolor{MyDarkBlue}{kaladharan.ajay@okstate.edu}},
{}~\textbf{and Shaikh Saad$^5$}\footnote{ E-mail: \textcolor{MyDarkBlue}{shaikh.saad@unibas.ch}}
}
\vspace{0.3cm}
{
\\\em $^1$Max-Planck-Institut f{\"u}r Kernphysik, Saupfercheckweg 1, 69117 Heidelberg, Germany
\\
$^2$Department of Physics, University of California, Berkeley, California 94720, USA
\\
$^3$Department of Physics, University of California, San Diego, California, 92093, USA
\\
$^4$Department of Physics, Oklahoma State University, Stillwater, OK 74078, USA
\\
$^5$Department of Physics, University of Basel, Klingelbergstrasse\ 82, \\ CH-4056 Basel, Switzerland
} 
\end{center}
\renewcommand{\thefootnote}{\arabic{footnote}}
\setcounter{footnote}{0}
\thispagestyle{empty}
%%%%%%%%%%%%%%%%%%%%%%%%%%%%%%%%%%%%%%%%%%%%%%%
%%%%%%%%%%%%%%%%%%%%%%%%%%%%%%%%%%%%%%%%%%%%%%%
%%%%%%%%%%%%%%%%%%%%%%%%%%%%%%%%%%%%%%%%%%%%%%%
%%%%%%%%%%%%%%%%%%%%%%%%%%%%%%%%%%%%%%%%%%%%%%%
\begin{abstract}
\noindent 
 We study the gravitational wave imprints of left-right symmetric model equipped with universal seesaw mechanism allowing for the natural generation of hierarchical masses of the Standard Model fermions. The scalar sector of this model is the minimal one, consisting of only two Higgs doublets. Following the construction of the full thermal potential for this model, we perform a scan of the entire parameter space and identify the region in which the cosmic phase transition associated with the left-right symmetry breaking gives gravitational wave signals detectable by a variety of planned space-based interferometers. Then we also discuss the relevant collider implications of this beyond the Standard Model scenario.
\end{abstract}

\newpage
\setcounter{footnote}{0}
{
  \hypersetup{linkcolor=black}
}
%%%%%%%%%%%%%%%%%%%%%%%%%%%%%%%%%%%%%%%%%%%%%%%%%%%%%%%%%%%%
%%%%%%%%%%%%%%%%%%%%%%%%%%%%%%%%%%%%%%%%%%%%%%%%%%%%%%%%%%%%
\section{Introduction}
The left-right symmetric models \cite{Pati:1974yy, Mohapatra:1974hk, Mohapatra:1974gc, Senjanovic:1975rk, Mohapatra:1980yp} represent a well-motivated gauge extension of the Standard Model (SM) providing a rich phenomenological framework and as such they have attracted a significant portion of attention over the years with scenarios incorporating a variety of possible scalar and fermion sectors. In the usual setup, the scalar sector is extended by a bidoublet field incorporating the SM Higgs and a pair of triplet scalars with the right-handed one being responsible for breaking of the left-right gauge symmetry, $SU(3)_C\otimes SU(2)_L\otimes SU(2)_R\otimes U(1)_{B-L}$, to the SM gauge group. The convenience of the left-right symmetric framework stems from several facts, such as the natural incorporation of right-handed neutrinos allowing for neutrino mass term and leading to the cancellation of the $B-L$ gauge anomaly. Restoring the symmetry between the right-handed and left-handed particles is also generally seen as an intermediate step \cite{Lindner:1996tf,Arbelaez:2013nga,Deppisch:2017xhv,Saad:2017pqj} of a grand unification \cite{Georgi:1974sy, Georgi:1974yf, Georgi:1974my, Fritzsch:1974nn} at high energies.

In this work, we consider a left-right symmetric framework with the simplest Higgs sector consisting of only two Higgs doublets, a left-handed and a right-handed under the two $SU(2)$ group factors. This leads to only two neutral physical Higgs states, one of them being the SM Higgs boson of mass 125 GeV.  Due to the minimality of the Higgs sector, vector-like fermions are introduced to generate masses of the SM fermions via universal seesaw mechanism \cite{Berezhiani:1983hm,Chang:1986bp,Davidson:1987mh,Rajpoot:1987fca,Babu:1988mw,Babu:1989rb, Jana:2021tlx}. One of the most attractive features of this class of models is, strong hierarchical structure of the SM fermions is naturally realized without the requirement of fine-tuning of the Yukawa couplings. Unlike the SM, where the Yukawa couplings lie in the range $10^{-6}$ - 1, a smaller range of $10^{-3}$ - 1 is sufficient to explain fermion mass hierarchy in this model. The observed mass hierarchies of the SM fermions are naturally explained via the heaviness of the vector-like fermions having inverse mass ordering. Another motivation of this framework is to solve the strong CP problem based on parity symmetry \cite{Babu:1988mw,Babu:1989rb} (for a recent study, see also Ref. \cite{deVries:2021pzl}), which does not require the implementation of Peccei-Quinn symmetry \cite{Peccei:1977hh}. In this work, we, however, do not impose parity symmetry. As will be shown, parity symmetric solution that demands the left and the right couplings to be identical (for example, $g_L=g_R$) fails to correspond to the strong first-order phase transition. From the phenomenological point of view, an advantage of this setup over the conventional left-right symmetric model that utilizes Higgs triplets is the possibility of incorporating lighter gauge boson $W_R$ that may address flavour anomalies, see for example Ref.~\cite{Babu:2018vrl}.

Realization of any gauge extension of the SM at high energies would mean the existence of an early Universe phase transition, as the extended gauge symmetry must be broken down to the SM gauge group. In case this phase transition is strongly of the first-order, it would lead to the production of gravitational waves. Although this signal would be typically too weak to be observed by the ground-based detectors~\cite{LIGOScientific:2016aoc}, a variety of planned space-based interferometers such as LISA~\cite{Caprini:2015zlo}, BBO~\cite{Corbin:2005ny} or DECIGO~\cite{Kudoh:2005as} may be sensitive enough to detect it. Given the fact that the minimal SM does not provide a cosmic phase transition of first-order~\cite{Kajantie:1995kf,Kajantie:1996mn,Aoki:2006we,Bhattacharya:2014ara}, similar observation could be a hint of new particle physics. In this context, gravitational waves generated by cosmic phase transitions associated with a variety of beyond the SM scenarios have been studied~\cite{Croon:2018erz,Okada:2018xdh,Baldes:2018nel,Chiang:2018gsn,Alves:2018oct,Baldes:2018emh,Ellis:2018mja,Madge:2018gfl,Ahriche:2018rao,Prokopec:2018tnq,Fujikura:2018duw,Beniwal:2018hyi,Brdar:2018num,Miura:2018dsy,Addazi:2018nzm,Shajiee:2018jdq,Marzo:2018nov,Breitbach:2018ddu,Angelescu:2018dkk,Alves:2018jsw,Kannike:2019wsn,Fairbairn:2019xog,Hasegawa:2019amx,Helmboldt:2019pan,Dev:2019njv,Bian:2019zpn,Mohamadnejad:2019vzg,Kannike:2019mzk,Bian:2019szo,Paul:2019pgt,Dunsky:2019upk,Athron:2019teq,Bian:2019kmg,Wang:2019pet,Alves:2019igs,DeCurtis:2019rxl,Addazi:2019dqt,Alanne:2019bsm,Huang:2020bbe,Alanne:2020jwx,Ertas:2021xeh,Huang:2020crf,Reichert:2021cvs}, including the conventional left-right symmetric setup~\cite{Brdar:2019fur}. The electroweak phase transition may become of first-order due to the presence of additional terms in the scalar potential,  which in the context of left-right symmetric model have been analysed in Refs.~\cite{Choi:1992wb,Barenboim:1998ib}. Here we study the possibility of generation of detectable gravitational waves within the phase transition associated with the aforementioned left-right symmetric model with minimal Higgs sector and universal seesaw mechanism providing the masses of the SM fermions. Indeed, we find that a similar gravitational wave signal can be produced during the left-right symmetry breaking in our model, with part of the model parameter space being within the reach of the future space-based detectors.

The following text is organized in this way: after we introduce the studied model in~Section~\ref{sec:2}, spelling out the considered particle content and its interactions, we investigate in detail the transition between the left-right symmetric phase and the SM phase in Section~\ref{sec:3}, identifying the region of the parameter space in which it is of the first-order. Consequently, in Section~\ref{sec:4}, we focus on the gravitational wave signal associated with this cosmic phase transition, and we confront it with sensitivities of the planned experiments. After discussing the possible collider signatures of our model in Section~\ref{col} we summarize and conclude in Section~\ref{sec:Con}.

%%%%%%%%%%%%%%%%%%%%%%%%%%%%%%%%%%%%%%%%%%%%%%%%%%%%%%%%%%%%
%%%%%%%%%%%%%%%%%%%%%%%%%%%%%%%%%%%%%%%%%%%%%%%%%%%%%%%%%%%%
\section{The Model}
\label{sec:2}
%%%%%%%%%%%%%%%%%%%%%%%%%%%%%%%%%%%%%%%%%%%%%%%%%%%%%%%%%%%%
%%%%%%%%%%%%%%%%%%%%%%%%%%%%%%%%%%%%%%%%%%%%%%%%%%%%%%%%%%%%
\subsection{Scalar sector}
The gauge group of the standard left-right symmetric model is $SU(3)_C\otimes SU(2)_L\otimes SU(2)_R\otimes U(1)_{B-L}$. This group, in our scenario, is spontaneously broken to the desired gauge symmetry at low energy in two steps, $SU(3)_C\otimes SU(2)_L\otimes SU(2)_R\otimes U(1)_{B-L} \rightarrow SU(3)_C\otimes SU(2)_L\otimes U(1)_{Y} \rightarrow SU(3)_C\otimes U(1)_{\mathrm{em}}$, via vacuum expectation values (VEVs) of two doublets (a left-handed and a right-handed) having the following quantum numbers
  \begin{eqnarray}
  \chi_L = \left(\begin{matrix}\chi_L^+ \\ \chi_L^0   \end{matrix}  \right)\sim (1,2,1,+1),~~~~  \chi_R = \left(\begin{matrix}\chi_R^+ \\ \chi_R^0   \end{matrix}  \right)\sim (1,1,2,+1).
    \end{eqnarray}
The scalar potential has the following simple form,
\begin{eqnarray}
V = -(\mu_L^2 \chi_L^\dagger \chi_L +\mu_R^2 \chi_R^\dagger \chi_R) +  \frac{\lambda_{L}}{2}(\chi_L^\dagger \chi_L)^2+ \frac{\lambda_{R}}{2}(\chi_R^\dagger \chi_R)^2   + \lambda (\chi_L^\dagger \chi_L)(\chi_R^\dagger \chi_R).
\label{V}
\end{eqnarray}
In the parity symmetric scenario,  $\lambda_{L}=\lambda_{R}$ and $\mu_L^2=\mu_R^2$ relations hold, where the latter can be violated by allowing for soft breaking of parity symmetry.  In this work, we focus on the parity asymmetric case for which none of these relations is valid, in general. 

After the gauge symmetry is spontaneously broken, associated gauge bosons become massive by eating-up $\chi^+_{L,R}$ and $\mathrm{Im}[\chi^0_{L,R}]$ scalar degrees of freedom. This leaves us with only two real physical scalars $\sigma_{L,R}\equiv \sqrt{2} \mathrm{Re}[\chi^0_{L,R}]$ and their corresponding zero temperature VEVs are  $\langle \sigma _L\rangle =v_L = 246.22\;  \textrm{GeV}$ and $\langle \sigma_R\rangle =v_R$, respectively. Without loss of generality, both these VEVs can be taken to be real. At the minimum of the potential, the following two stationary conditions must be satisfied
\begin{align}
-\mu^2_{L(R)}+\frac{1}{2}\lambda_{L(R)}v^2_{L(R)}+\frac{1}{2}\lambda v^2_{R(L)}=0.    
\end{align}

The mass-squared matrix for the  physical scalars takes the form:
\begin{eqnarray}
{\cal M}^2_{\sigma} = \left[\begin{matrix} -\mu^2_{L}+\frac {3 \lambda_{L}}{2} v_L^2+\frac {\lambda}{2} v_R^2 &  \lambda v_L v_R \\  \lambda v_L v_R & -\mu^2_{R}+\frac {3 \lambda_{R}}{2} v_R^2+\frac {\lambda}{2} v_L^2   \end{matrix}\right],
\end{eqnarray}
which leads to two non-zero eigenvalues. We identify the smaller mass eigenstate (of mass $M_h$) as the SM Higgs $h$, while $H$ denotes the heavier mass eigenstate (of mass $M_H$).  The mixing of these two states is parameterized by, 
\begin{equation}
    \tan \xi=\frac {2\lambda v_Lv_R}{\lambda_{R}v_R^2-\lambda_{L}v_L^2}.
\end{equation}
Furthermore, the masses of the Goldstone bosons are given by,
\begin{align}
    &M_{1,2}^2=-\mu^2_{L,R}+\frac {\lambda_{L,R}}{2} v_{L,R}^2+\frac {\lambda}{2} v_{R,L}^2.
\end{align}
For the convenience of our numerical analysis, we utilize the aforementioned stationary conditions and treat the tree-level masses $M_{h,H}$, zero temperature VEV $v_R$, and the mixing angle $\xi$ as free parameters. Then the couplings appearing in the scalar potential can be expressed in terms of these independent quantities as,
\begin{align}
    &\lambda_{L,R}=\frac {1}{v_{L,R}^2}\left ( M^2_{h,H}\cos^2\xi + M^2_{H,h}\sin^2\xi \right ),\;\;
    \lambda=\frac {\sin 2\xi}{2v_Lv_R}\left (M^2_H-M^2_h \right ).\label{lamR}
\end{align}
In our numerical procedure, we guarantee the perturbativity of these couplings by demanding $\lambda_i \leq \sqrt{4\pi}$, and the boundedness of the potential requires:
\begin{equation}
\lambda_{L} \geq 0, ~~~\lambda_{R} \geq 0,~~~\lambda \geq -\sqrt{\lambda_{L}\lambda_{R}}~.
\end{equation}

%%%%%%%%%%%%%%%%%%%%%%%%%%%%%%%%%%%%%%%%%%%%%%%%%%%%%%%%%%%%
%%%%%%%%%%%%%%%%%%%%%%%%%%%%%%%%%%%%%%%%%%%%%%%%%%%%%%%%%%%%
\subsection{Gauge sector}
In this model, in addition to the SM charged vector bosons $W_L^\pm$, their right-handed partners $W_R^\pm$ also exist. They, however, do not mix at the tree-level and their masses can be straightforwardly computed as
\begin{equation}
 M^2_{W^\pm_L}~=~\frac{g^2_Lv^2_L }{4},~~~ M^2_{W^\pm_R}~=~\frac{g^2_Rv^2_R }{4}~.
\end{equation}
The neutral vector boson sector is more complicated due to the tree-level mixings. We denote these mixed states as $(W_{3L},\,W_{3R},\,B)$, where $B$ represents the $B-L$ vector boson. As for the mass eigenstates,  the photon field $A_\mu$ remains massless, as it should,  while the two orthogonal fields $Z_L$ and $Z_R$ have the following mass-squared matrix:
\begin{eqnarray}
{\cal M}^2_{Z_L-Z_R} =
\frac{1}{4} \, \left( \begin{matrix} (g_Y^2+g_L^2)\, v_L^2 & g_Y^2 \sqrt{\frac{g_Y^2 + g_L^2}{g_R^2-g_Y^2}} \,v_L^2 \\
 g_Y^2 \sqrt{\frac{g_Y^2 + g_L^2}{g_R^2-g_Y^2}}\, v_L^2 & \frac{g_R^4}{g_R^2-g_Y^2}\, v_R^2 + \frac{g_Y^4}{g_R^2-g_Y^2}\, v_L^2      \end{matrix} \right).
 \label{ZLR}
\end{eqnarray}
Since the mixing angle between these two states, $\theta_Z\sim (g^2_Y/g^4_R) (v^2_L/v^2_R) \sqrt{(g^2_L+g^2_Y)(g^2_R-g^2_Y)}$, is tiny for all practical purposes of our analysis, the mass eigenstates are very well approximated with their corresponding gauge eigenstates. Consequently, the contribution to the electroweak precision T-parameter is negligible.

Gauge coupling strengths of the $SU(2)_R$ and $U(1)_{B-L}$ groups are labeled by $g_R$ and $g_{B}$, and the coupling $g_Y$ associated with the hypercharge group $U(1)_Y$ is defined in terms of $g_R$ and $g_{B}$. This embedding of $U(1)_Y\subset SU(2)_R \otimes U(1)_{B-L}$ leads to the following matching condition for $g_Y$:
\begin{equation}
\frac{Y}{2} = T_{3R} + \frac{B-L}{2} ~~~~~ \Rightarrow~~~~~ \frac{1}{g_Y^{2}} = \frac{1}{g_R^{2}} +\frac{1}{g_B^{2}}.
\label{embed}
\end{equation} 
We take $g^2_Y=0.1279$ (here we have used $\alpha(M_Z)=1/127.9$ and $\sin^2\theta_W(M_Z)=0.2315$) and the consistency of the above equation requires $g^2_R \geq g^2_Y$. In our numerical scan, we vary $g_R$ within the range $[g_Y,\sqrt{4\pi}]$ and  $g_B$ is determined accordingly using the known value of $g_Y$. 

%%%%%%%%%%%%%%%%%%%%%%%%%%%%%%%%%%%%%%%%%%%%%%%%%%%%%%%%%%%%
%%%%%%%%%%%%%%%%%%%%%%%%%%%%%%%%%%%%%%%%%%%%%%%%%%%%%%%%%%%%
\subsection{Fermion sector}
The SM chiral fermions in this theory are defined in the usual way,
\begin{align}
&Q_{L}=\begin{pmatrix}u_L\\d_L\end{pmatrix}\sim \left(3,2,1,+\frac{1}{3}\right),\;\; Q_{R}=\begin{pmatrix}u_R\\d_R\end{pmatrix}\sim \left(3,1,2,+\frac{1}{3}\right),
\\
&L_{L}=\begin{pmatrix}\nu_L\\e_L\end{pmatrix}\sim \left(1,2,1,-1\right),\;\; L_{R}=\begin{pmatrix}\nu_R\\e_R\end{pmatrix}\sim \left(3,1,2,-1\right), 
\end{align}
where, we have suppressed the group as well as family indices (as usual $i,j=1-3$). Note that the above set of chiral fermions automatically contains the right-handed neutrino due to the left-right symmetry. The scalar sector employed to break the gauge symmetry, however, is unable to provide masses to these fermions. This is because scalar doublets cannot couple to any fermion bilinears constructed from the aforementioned set of chiral fermions. Hence, to recover the masses of the SM fermions, we introduce vector-like fermions (three families with $I,J=1-3$) in the following representations
\begin{align}
&U\sim \left(3,1,1,\frac{4}{3}\right),\;\;   D\sim \left(3,1,1,-\frac{2}{3}\right),\;\;   E\sim \left(1,1,1,-2\right),\;\;   N\sim \left(1,1,1,0\right).     
\end{align}
With these additional fermions, SM fermion masses arise via the seesaw mechanism, which is commonly referred to as the universal seesaw mechanism \cite{Berezhiani:1983hm,Chang:1986bp,Davidson:1987mh,Rajpoot:1987fca,Babu:1988mw,Babu:1989rb}. The Yukawa part of the Lagrangian with the addition of the vector-like states takes the following form:
\begin{align}
\mathcal{L}_Y&= Y^{L(R)}_U\overline Q_{L(R)}\widetilde \chi_{L(R)} U_{R(L)} + Y^{L(R)}_D\overline Q_{L(R)} \chi_{L(R)} D_{R(L)} + Y^{L(R)}_E\overline L_{L(R)} \chi_{L(R)} E_{R(L)}
\nonumber \\&
+M_{U}\overline U_LU_R+M_{D}\overline D_LD_R+M_{E}\overline E_LE_R+h.c.~,
\end{align}
where $\widetilde \chi= \epsilon \chi^\ast$ is used. Then the $6\times 6$ mass matrix in each sector shares the same form,
\begin{align}
\mathcal{M}_{U,D,E}=\begin{pmatrix}0&\frac{v_L}{\sqrt{2}} Y^L_{U,D,E}\\
\frac{v_R}{\sqrt{2}} {Y^{R \dagger}_{U,D,E}}&M_{U,D,E}\end{pmatrix}.    
\end{align}
We write the above-defined matrix $\mathcal{M}_{U}$ in $\left(u, c, t, U, C, T\right)\equiv \left(f, F\right)$ basis (and similarly in the down-quark and charged lepton sectors). An attractive feature of this model is the fact that the strong hierarchical mass spectrum of the SM fermions can be explained using the seesaw structure. Without loss of generality, one can work in a basis where $M_{U,D,E}$ are simultaneously diagonal, which we adopt in the following. Ignoring generation mixing, the mass of a light state $f$ is given by $m_f\sim \frac{1}{2}v_Lv_R Y^L_FY^R_F M^{-1}_F$. Assuming Yukawa couplings of the same order, this clearly implies an inverse mass ordering for the heavy states. Note that to get the correct top-quark mass, the associated Yukawa couplings are required to be of order unity, whereas the rest of the couplings can be taken to be somewhat smaller, which is the scenario we focus on. Then, only the top-quark partner will contribute to the effective potential (see the next section). Its mass is given by $m_T^2=M^2_T+v_R^2{Y^R_{T}}^2/2$. For the rest of our analysis we fix $Y_T^R=1$ and $M_T=v_R$ (as before $v_R$ represents the zero temperature VEV). 

On the other hand, the Yukawa Lagrangian associated with the neutral fermion sector  contains the following terms 
\begin{align}
\mathcal{L}^\nu_Y&= Y^{L(R)}_\nu\overline L_L \widetilde \chi_{L(R)} N_{R(L)}  + \hat Y^{L(R)}_\nu\overline L_{L(R)} \widetilde \chi_{L(R)} N^c_{R(L)}   
+M_N \overline N_L N_R +M_{L(R)} \overline N^c_{L(R)} N_{L(R)} +h.c..
\end{align}
Here,  $M_{L(R)}$ and $M_N$ are Majorana and Dirac mass terms, respectively. From the above Lagrangian, the $12\times 12$ mass matrix in the basis $\left(\nu, \nu^c, N, N^c\right)_L$ can be written as
\begin{align}
\mathcal{M}_\nu=
\begin{pmatrix}
0&0&\frac{v_L}{\sqrt{2}}Y^{L}_\nu&\frac{v_L}{\sqrt{2}}\hat Y^{L}_\nu\\
0&0&\frac{v_R}{\sqrt{2}}Y^{R}_\nu&\frac{v_R}{\sqrt{2}}\hat Y^{R}_\nu\\
\frac{v_L}{\sqrt{2}} Y^{L T}_\nu&\frac{v_R}{\sqrt{2}} Y^{R T}_\nu&M_L&M_N\\
\frac{v_L}{\sqrt{2}} \hat Y^{L T}_\nu&\frac{v_R}{\sqrt{2}} \hat Y^{R T}_\nu&M^T_N&M_R
\end{pmatrix},
\end{align}
where the $M_{L,R}$ matrices are symmetric due to their Majorana nature. A variety of different mass spectra can emerge from the above neutrino mass matrix. For instance, assuming the bare mass parameters of similar order, $M_L\sim M_R\sim M_N =M$, masses of $N, N^c$ will be of the same order, $\sim M$, whereas $\nu$ and $\nu^c$ will receive masses of order of $v^2_L/M$ and $v^2_R/M$, respectively (we have assumed all Yukawa couplings to be of order unity). Then, a choice of large $M$ ($M\gg v_R$) naturally explains the observed tiny masses of the active neutrinos without requiring small Yukawa couplings. Subsequently, the masses of $\nu^c$ are expected to be around the MeV range. On the other hand, assuming $M\sim v_R$ would result in $\nu^c, N, N^c$ all having masses of order of $v_R$. In such case, the corresponding Yukawa couplings are required to be small (depending on the scale of $v_R$) to recover the light neutrino masses.   
 
%%%%%%%%%%%%%%%%%%%%%%%%%%%%%%%%%%%%%%%%%%%%%%%%%%%%%%%%%%%%
%%%%%%%%%%%%%%%%%%%%%%%%%%%%%%%%%%%%%%%%%%%%%%%%%%%%%%%%%%%%
\subsection{Effective Potential} \label{EP}
In this section, we discuss the effective potential \cite{Heisenberg:1936nmg,Schwinger:1951nm,Goldstone:1962es,Jona-Lasinio:1964zvf,Coleman:1973jx} at finite temperature within the left-right symmetric model. Since experimental measurements require the left-right symmetry breaking scale to be much higher than the EW breaking scale, $v_R\gg v_L$, it is enough, to a very good approximation, to consider only $\sigma_R$ dependent effective potential to investigate the high energy phase transition. From henceforth, we denote the temperature dependent VEV of the field $\sigma_R$ by $r$. The one-loop, daisy-improved finite-temperature effective potential takes the form
\begin{align}
V_{\mathrm{eff}}(r, T)= V_{\mathrm{tree}} + V_{\mathrm{CW}}+V_{\mathrm{T}}, 
\label{VEFF} 
\end{align}
where $V_{\mathrm{CW}}$ is the zero-temperature Coleman-Weinberg (CW) contribution and $V_T$ contains finite temperature terms. The one-loop CW potential, renormalized in the $\overline{\textrm{MS}}$ scheme  is given by \cite{Coleman:1973jx}
\begin{align}
V_{\mathrm{CW}}(r)=\frac{1}{64\pi^2}\sum_{i}n_i\left(-1\right)^{2s_i} M^4_i(r)\left(  \log\left[\frac{M^2_i(r)}{\Lambda^2}\right] - c_i \right).
\end{align}
Here, $M_i(r)$ are the background dependent masses of the particles in our model (which are obtained by setting $v_L=0$ and $v_R=r$ in all the expressions for the masses given in the previous subsections) and the summation runs over all the assumed fermionic and bosonic states (including Goldstones). Here, $\Lambda$ represents the renormalization scale, which we fix to be $v_R$, and the CW potential is evaluated in the Landau gauge~\cite{Quiros:1999jp}; hence, there are no ghost contributions. Although we are working in the Landau gauge, the mass eigenvalues of the Goldstones can be non-zero even for $T=0$, as these are evaluated at field configurations rather than the tree-level VEVs at zero temperature. The spin and the number of degrees of freedom (dof) of a particle $i$ is denoted by $s_i$ and $n_i$, respectively. Moreover, in the $\overline{\textrm{MS}}$ scheme: $c_i=3/2 (5/6)$ for scalars and fermions (vector bosons).

The finite temperature one-loop quantum correction is given by~\cite{Dolan:1973qd, Quiros:1999jp},
\begin{align}
V_{\mathrm{T}}(r, T)=\frac{T^4}{2\pi^2}\sum_{i}n_i J_i\left( \frac{M^2_i(r)}{T^2} \right),
\end{align}
where the summation is taken over all scalars, fermions, and vector bosons (both transverse and longitudinal components). The thermal functions are defined by the following integrals~\cite{Dolan:1973qd, Parwani:1991gq,Arnold:1992rz, Quiros:1999jp,Carena:2008vj},
\begin{align}
J_{B,F}(y^2)=\int^{\infty}_0 \mathrm{d}x\; x^2 \log \left( 1\mp e^{-\sqrt{x^2+y^2}} \right). 
\end{align}
In the thermal correction we have included also the daisy resummation~\cite{Carrington:1991hz}, which amounts to the following replacement 
\begin{align}
J_i\left( \frac{M^2_i(r)}{T^2} \right)\to J_i\left( \frac{M^2_i(r)}{T^2} \right) - \frac{\pi}{6 T^3} \left( \{ M^2_i(r) +\Pi_i(r,T) \}^{3/2} - \{ M^2_i(r) \}^{3/2} \right)
\end{align}
for the scalar fields and the longitudinal components of the gauge bosons. In our model the associated self-energy corrections are given by~\cite{Comelli:1996vm},
\begin{align}
&\Pi_{\chi_L}(r,T)= \left [ \frac {\lambda_{L}}{4}+\frac {\lambda}{6}+\frac {3g_{L}^2+g_{BL}}{16} \right ]T^2,\\
&\Pi_{\chi_R}(r,T)= \left [ \frac {\lambda_{R}}{4}+\frac {\lambda}{6}+\frac {3g_{R}^2+g_{BL}^2}{16}+\frac{1}{4} {Y_T^R}^2 \right ]T^2,
\\&
\Pi^L_{W_R}(r,T)=\frac{11}{6} g^2_RT^2,\;\;
\Pi^L_{Z_L,Z_R,Z_{BL}}(r,T)=T^2\bigg\{ \frac{11}{6}g_L^2, \frac{11}{6}g_R^2, \frac{44}{3}g^2_{B-L}\bigg\}.
\end{align}

Note that even at zero temperature the masses and mixings arising from the effective potential differ from their expected tree-level values. Masses computed in this way correspond to the full one-loop corrected masses in the approximation of
zero external momenta. Subsequently, these loop-corrected masses should be used in determining the coupling parameters of the theory that makes the procedure numerically expensive. However, for numerical efficiency, we adopt the modified renormalization scheme for the CW potential as described in Ref.~\cite{Basler:2016obg} (for more details see also Refs.~
\cite{Casas:1994us, Cline:1996mga, Fromme:2006cm, Cline:2011mm, Dorsch:2013wja, Dorsch:2014qja, Martin:2014bca, Elias-Miro:2014pca,  Camargo-Molina:2016moz, Krause:2016oke, Krause:2016xku}). Following their prescriptions, we add relevant counterterm potential that has the form,
\begin{equation}
    V_{\mathrm{CT}}(h,r)=-\left(\frac {\delta \mu ^2_L}{2}h^2+\frac {\delta  \mu ^2_R}{2}r^2\right)+\frac {\delta  \lambda_{L}}{4}h^4+\frac {\delta  \lambda_{R}}{4}r^4+\frac {\delta  \lambda}{4}h^2r^2. \label{VCT}
\end{equation}
Here, we have for generality kept contributions coming from both the scalar fields. To find the global minimum at a given temperature $T$, the effective potential of Eq.~\eqref{VEFF} including the counterterm potential Eq.~\eqref{VCT} must be minimized. The coefficients of the counterterms are then obtained via the following five conditions,
\begin{align}
    \frac{\partial V_{\mathrm{CT}}(h,r)}{\partial x_i}\bigg\rvert_{(h=v_L,r=v_R)}&=-\frac{\partial V_{\mathrm{CW}}(h,r)}{\partial x_i}\bigg\rvert_{(h=v_L,r=v_R)}, 
    \\
    \frac{\partial^2 V_{\mathrm{CT}}(h,r)}{\partial x_i \partial x_j}\bigg\rvert_{(h=v_L,r=v_R)}&=-\frac{\partial^2 V_{\mathrm{CW}}(h,r)}{\partial x_i \partial x_j}\bigg\rvert_{(h=v_L,r=v_R)},
\end{align}
for $x_i\in \{h,r\}$.
Partial derivatives of CW potential are computed consistently in the Landau gauge using the recipe provided in Ref. \cite{Camargo-Molina:2016moz}.

As will be discussed later, the requirement of a strong first-order phase transition that corresponds to $\frac {r_c}{T_c} \geq 1$ (where $T_c$ is the critical temperature and $r_c$ is the VEV at $T_c$) is fairly restrictive in our model due to the very simple structure of the scalar sector. Specifically, the strong first-order phase transition demands somewhat large gauge coupling $g_R$ and small values of the quartic coupling $\lambda_R\ll 1$.  As a result, the one-loop contributions in the CW effective potential arising from the gauge boson mediated processes start to
dominate over the quartic coupling contribution and may cause the broken vacuum  $V\left(\langle \chi_R\rangle\neq 0 \right)$ to become unstable. To achieve a consistent symmetry breaking scenario, we satisfy the Linde-Weinberg bound \cite{Linde:1975sw, Weinberg:1976pe} for each parameter set by imposing the following condition~\cite{Politzer:1978ic,Basecq:1988cv} 
\begin{equation}
    V_{\mathrm{eff}}(v_R,0)<V_{\mathrm{eff}}(0,0).
    \label{Eq:LW1}
\end{equation}

Before closing this section, we would like to make a remark that since the properties of the SM Higgs are very well measured in the experiments that suggest almost no deviation from the SM predictions and require small mixing with exotic scalars, we set for simplicity the mixing angle $\xi$ (or equivalently the coupling $\lambda$) to zero. Consequently, the measured value of the SM Higgs mass $M_h=125$ GeV fixes the coupling $\lambda_L$ uniquely. As a result, with fixed $Y^R_T=1$ as mentioned before, we are left with only two coupling parameters $\{\lambda_R, g_R\}$ in this theory that determine the behaviour of the symmetry breaking at high energy scale. We vary $v_R$ and $g_R$ within the ranges $v_R~\in~[1, 20]$~TeV and $g_R~\in~[g_Y, \sqrt{4\pi}]$, whereas $M_H$ is varied from 300 GeV to a maximum value such that the condition $\lambda_R \leq \sqrt{4\pi}$ is assured to be satisfied (see Eq.~\ref{lamR}).

%%%%%%%%%%%%%%%%%%%%%%%%%%%%%%%%%%%%%%%%%%%%%%%%%%%%%%%%%%%%
%%%%%%%%%%%%%%%%%%%%%%%%%%%%%%%%%%%%%%%%%%%%%%%%%%%%%%%%%%%%
\section{Phase Transition Analysis}
\label{sec:3}
%%%%%%%%%%%%%%%%%%%%%%%%%%%%%%%%%%%%%%%%%%%%%%%%%%%%%%%%%%%%
%%%%%%%%%%%%%%%%%%%%%%%%%%%%%%%%%%%%%%%%%%%%%%%%%%%%%%%%%%%%
As the focus of this work is to study the potential gravitational wave signature arising from the breaking of the left-right symmetry in the above described model, we are primarily interested in identification of the subset of the parameter space for which this phase transition is strongly of first-order. Generally, a first-order phase transition is characterized by critical temperature $T_c$, at which the true vacuum and the false vacuum are degenerate, and nucleation temperature $T_n$, at which the transition actually occurs. The probability of tunneling from the false vacuum to the true one at a temperature below $T_c$ is given by~\cite{Linde:1980tt,Coleman:1977py},
\begin{equation}
    \Gamma (T)\approx T^4\left (\frac {S_3}{2\pi T}  \right )^{3/2}e^{-\frac {S_3}{T}},
\end{equation}
where $S_3$ is the three-dimensional Euclidean action corresponding to the critical bubble. We can calculate $S_3$ using
\begin{equation}
    S_3=\int_{0}^{\infty}{\mathrm{d}r\mathrm{d}r^2\left [ \frac 12\left ( \frac {\mathrm{d}\phi(r)}{\mathrm{d}r} \right )^2+V(\phi,T) \right ]}.
\end{equation}
The scalar field $\phi$ is obtained by solving scalar field's equation of motion
\begin{equation}
    \frac {\mathrm{d}^2\phi}{\mathrm{d}r^2}+\frac 2r \frac{\mathrm{d}\phi}{\mathrm{d}r}=\frac {\mathrm{d}V(\phi,T)}{\mathrm{d}r},
\end{equation}
subjected to boundary condition $\lim\limits_{r\rightarrow \infty}\phi(r)=0$ and $\lim\limits_{r\rightarrow 0}\frac {\mathrm{d}\phi(r)}{\mathrm{d}r}=0$. We use the CosmoTransitions package~\cite{Wainwright:2011kj} to solve the differential equation and compute the action $S_3$. The phase transition occurs at nucleation temperature $T_n$, which is defined as the temperature for which on average one bubble nucleates per horizon volume~\cite{Moreno:1998bq},
\begin{equation} 
\label{equ:Tncond}
    \int_{T_n}^{\infty}\frac {\mathrm{d}T}{T}\frac {\Gamma (T)}{H(T)^4}=1.
\end{equation}
At this temperature, the phenomenological quantities $\alpha$ and $\beta$ are defined. The parameter $\alpha$ measures the strength of the phase transition and corresponds to vacuum energy released during the phase transition normalized by total radiation energy density~\cite{Espinosa:2010hh},
\begin{equation}
    \alpha=\frac{\rho_{\mathrm{vac}}}{\rho_{\mathrm{rad}}}=\frac{1}{\rho_{\mathrm{rad}}}\left [\frac T4 \frac {\mathrm{d}\Delta V}{\mathrm{d}T}-\Delta V \right ]_{T_n}.
\end{equation}
Here, radiation energy density $\rho_{\mathrm{rad}}$ is given by $\rho_{\mathrm{rad}}=\frac {g_{\star}\pi ^2T^4}{30}$, where $g_{\star}=132.5$ for our case. The parameter $\beta$ corresponds to inverse time duration of the phase transition~\cite{Kamionkowski:1993fg}, namely,
\begin{equation}
    \beta=\left (HT\frac {\mathrm{d}(S_3/T)}{\mathrm{d}T}  \right )_{T_n},
\end{equation}
where $H$ is the Hubble's rate.
\begin{figure}[th!]
  \centering
  \begin{minipage}[b]{0.65\textwidth}
    \includegraphics[width=\textwidth]{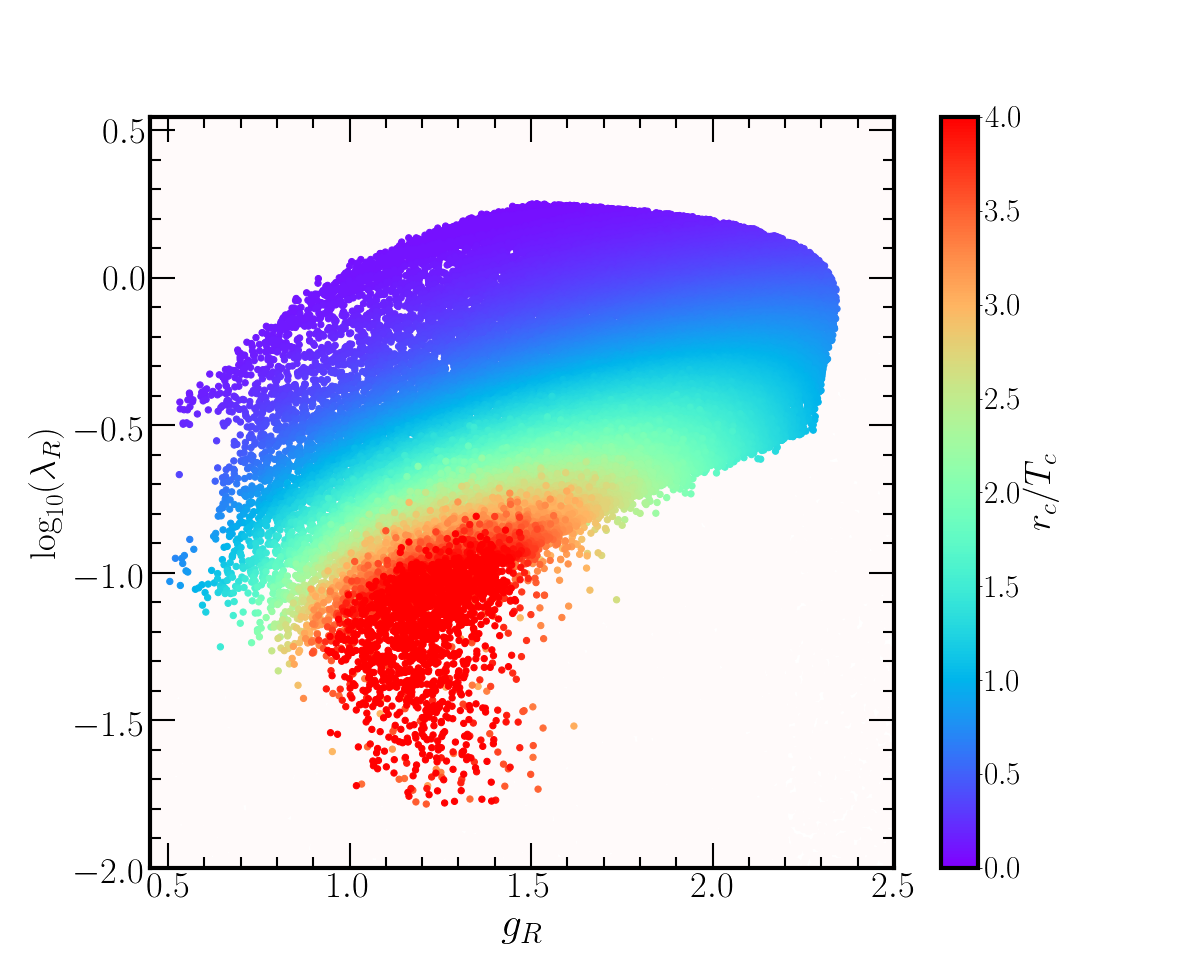}
  \end{minipage}
  \caption{Gauge coupling $g_R$ versus log of quartic coupling $\lambda_{R}$ colour-coded with strength of the left-right symmetric phase transition measured by $r_c/T_c$. Large $r_c/T_c$ values prefer small values of $\lambda_{R}$.  }
  \label{fig:ptstrength}
\end{figure}

As argued in the previous section, the left-right symmetry breaking in our model can be characterized by the pair of parameters $\{\lambda_R, g_R\}$. Therefore, we studied the phase transition as a function of these two couplings and the results of our numerical analysis are summarized in Fig.~\ref{fig:ptstrength}. There we show the identified parameter space allowing for the phase transition to be first-order with its strength represented by the colour of the points. Given the position of the red region associated with the strong first-order phase transition, it is apparent that scenarios with small values of $\lambda_R$ are most promising for the production of a detectable gravitational wave signal, which is consistent with findings of Ref.~\cite{Brdar:2019fur}. The associated values of the gauge coupling $g_R$ take on values between $1$ and $1.5$.  Preference for larger values of $g_R$ and smaller values of $\lambda_R$ rules out a large portion of the parameter space corresponding to $g_R>1.5$ and $\lambda_R<0.1$ (approximately the lower right half of the plot in  Fig.~\ref{fig:ptstrength}). This occurs because the associated vacuum becomes unstable in this region and fails to satisfy the Linder-Weinberg bound (for details, see Sec. \ref{EP}).  Besides the general study of the promising parameter space we selected five different successful benchmarks points, for which we show the relevant quantities characterizing the phase transition in Tab.~\ref{tab:1}. 

\begin{table}
\centering
\begin{tabular}{ c p{2.0cm}  p{2.0cm} p{2.0cm} p{2.0cm} p{2.0cm} p{2.0cm} }
\toprule
\toprule
& BP1  &BP2 & BP3 & BP4 & BP5 \\
\cmidrule{2-6}
$v_R$ [TeV]& $15.4374$& $17.3036$&  $8.66765$& $8.49708$& $6.45567$\\
$M_H$ [TeV]& $1.73089$& $2.62174$&  $0.33180$& $1.39295$& $1.46009$ \\
$g_R$& $1.32586$& $1.31839$& $1.08493$& $1.46333$& $1.73601$ \\
$M_{W_R}$[TeV]& $10.2340$& $11.4064$& $4.70190$& $6.21703$& $5.60356$\\
$M_{Z_R}$[TeV]& $10.6097$& $11.8302$& $4.96700$& $6.40254$& $5.72083$\\
$T_c$ [TeV]& $4.48375$& $4.95940$& $1.76057$& $2.97584$& $3.26440$\\
$T_n$ [TeV]& $1.12847$& $1.39340$& $0.56453$& $1.49604$& $2.10817$\\
$\alpha$& $0.90463$& $0.60801$& $0.52221$& $0.05074$& $0.01208$  \\
$\beta/{H_\star}$& $38.7374$& $72.2803$& $361.586$& $193.882$& $216.339$\\
\bottomrule
\bottomrule
\end{tabular}
\caption{Values of parameters characterizing the five selected benchmark points, for which we show the stochastic gravitational wave signal in Fig.~\ref{fig:gwbenchmark}. While $v_R$, $M_H$, and $g_R$ are free parameters varied in our numerical analysis, the rest are derived quantities. As explained in the text, we set $(\xi, Y^R_T)$= (0, 1), and $\lambda_{R}$ is determined using Eq.~\ref{lamR}.    }
\label{tab:1} 
\end{table}

%%%%%%%%%%%%%%%%%%%%%%%%%%%%%%%%%%%%%%%%%%%%%%%%%%%%%%%%%%%%
%%%%%%%%%%%%%%%%%%%%%%%%%%%%%%%%%%%%%%%%%%%%%%%%%%%%%%%%%%%%
\section{Gravitational Wave Signatures}
\label{sec:4}
%%%%%%%%%%%%%%%%%%%%%%%%%%%%%%%%%%%%%%%%%%%%%%%%%%%%%%%%%%%%
%%%%%%%%%%%%%%%%%%%%%%%%%%%%%%%%%%%%%%%%%%%%%%%%%%%%%%%%%%%%
First-order phase transition in the early Universe could generate gravitational wave signals observable today. These signals would peak around the millihertz region and they could be detected by the next-generation space-based detectors such as LISA \cite{Caprini:2015zlo}, BBO \cite{Corbin:2005ny} and DECIGO \cite{Kudoh:2005as}. There are three different sources of gravitational waves produced in the first-order phase transitions: bubble wall collisions, sound waves and magnetohydrodynamic turbulence in the plasma, i.e., the total gravitational wave strength is given by the sum of these as
\begin{equation}
    \Omega_{\mathrm{GW}}h^2=\Omega_{\mathrm{sw}}h^2+\Omega_{\mathrm{turb}}h^2+\Omega_{\mathrm{coll}}h^2.
\end{equation}

The contribution from sound waves can be modelled by~\cite{Hindmarsh:2013xza, Hindmarsh:2016lnk, Hindmarsh:2017gnf} the expression
\begin{equation}
    \Omega_{\mathrm{sw}}h^2=\frac {2.65 \times 10^{-6}}{\beta/H}\left ( \frac {k_{\mathrm{s}} \alpha}{1+\alpha} \right )^2\left ( \frac {100}{g_{*}} \right )^{1/3}v_{\mathrm{w}} \left ( \frac {f}{f_{\mathrm{sw}}} \right )^3\left ( \frac {7}{4+3(f/f_{\mathrm{sw}})^2} \right )^{7/2}\Upsilon (\tau_{\mathrm{sw}}).
\end{equation}
Here, we fix the bubble wall velocity to be $v_{\mathrm{w}}=1$ for our analysis and $k_{\mathrm{s}}$ denotes the efficiency factor for the sound wave contribution defined as~\cite{Caprini:2015zlo}
\begin{equation}
    k_{\mathrm{s}}=\frac {\alpha}{0.73+0.083\sqrt{\alpha}+\alpha}.
\end{equation}
Further, $f_{\mathrm{sw}}$ denotes the peak frequency of the sound wave contribution~\cite{Huber:2008hg},
\begin{equation}
    f_{\mathrm{sw}}=\frac {1.9\times 10^{-5}}{v_{\mathrm{w}}}\left ( \frac {\beta}{H} \right )\left ( \frac {T_n}{100GeV} \right )\left ( \frac {g_{*}}{100} \right )^{1/6} \mathrm{Hz}.
\end{equation}
The suppression factor,
\begin{equation}
    \Upsilon (\tau_{\mathrm{sw}})=1-\frac {1}{\sqrt{1+2\tau_{\mathrm{sw}} H_{\star}}},
\end{equation}
arises due to the finite lifetime $\tau_{\mathrm{sw}}$ of the sound waves~\cite{Guo:2020grp, Hindmarsh:2020hop} given by
~\cite{Hindmarsh:2017gnf},
\begin{equation}
    \tau_{\mathrm{sw}}=\frac {R_{\star}}{\overline{U_f}},
\end{equation}
where $R_{\star}\simeq (8\pi)^{1/3}v_{\mathrm{w}}/\beta$ is the mean bubble separation and $\overline{U_f}^2$ is the mean square velocity~\cite{Bodeker:2017cim} that can be obtained as
\begin{equation}
    {\overline{U_f}}^2=\frac 34 \frac {\alpha}{1+\alpha}k_{\mathrm{s}}.
\end{equation}

The gravitational wave spectrum from the magnetohydrodynamic turbulence can be parametrized as~\cite{Caprini:2009yp},
\begin{equation}
    \Omega_{\mathrm{turb}}h^2=\frac {3.35 \times 10^{-4}}{\beta/H}\left ( \frac {k_{\mathrm{t}} \alpha}{1+\alpha} \right )^{3/2}\left ( \frac {100}{g_{*}} \right )^{1/3}v_{\mathrm{w}} \left ( \frac {f}{f_{\mathrm{sw}}} \right )^3\frac {1}{\left [1+(f/f_{\mathrm{turb}})  \right ]^{11/3}(1+8\pi f/h_{*})},
\end{equation}
where we assume $k_{\mathrm{t}}=0.05k_{\mathrm{s}}$ based on numerical simulation~\cite{Caprini:2015zlo} and the peak frequency is given by~\cite{Caprini:2009yp},
\begin{equation}
    f_{\mathrm{turb}}=\frac {2.7\times 10^{-5}}{v_{\mathrm{w}}}\left ( \frac {\beta}{H} \right )\left ( \frac {T_n}{100GeV} \right )\left ( \frac {g_{*}}{100} \right )^{1/6} \mathrm{Hz},
\end{equation}
with 
\begin{equation}
    h_{*}=16.5\times 10^{-6}\left ( \frac {T_n}{100GeV} \right )\left ( \frac {g_{*}}{100} \right )^{1/6} \mathrm{Hz}.
\end{equation}

\begin{figure}[th!]
  \centering
  \begin{minipage}[b]{0.495\textwidth}
    \includegraphics[width=\textwidth]{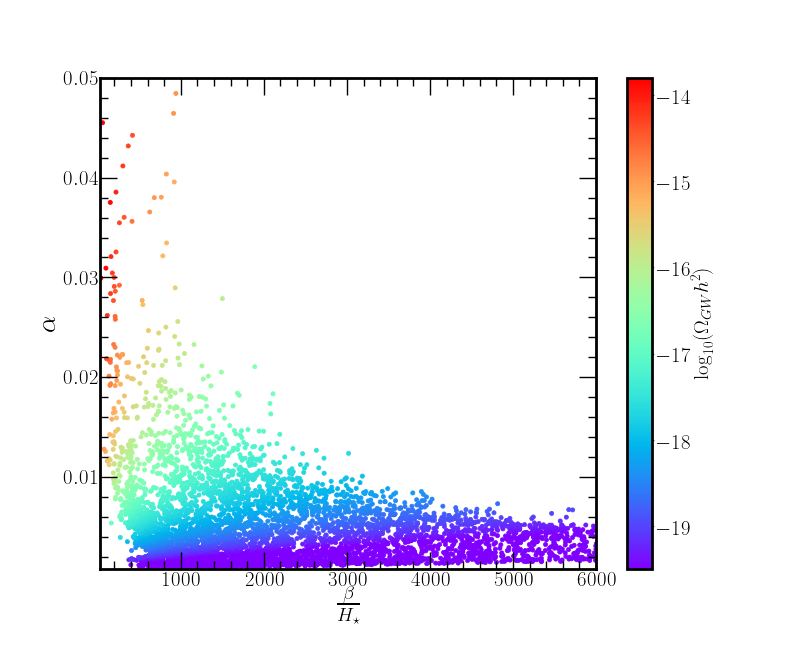}
  \end{minipage}
  \hfill
  \begin{minipage}[b]{0.495\textwidth}
    \includegraphics[width=\textwidth]{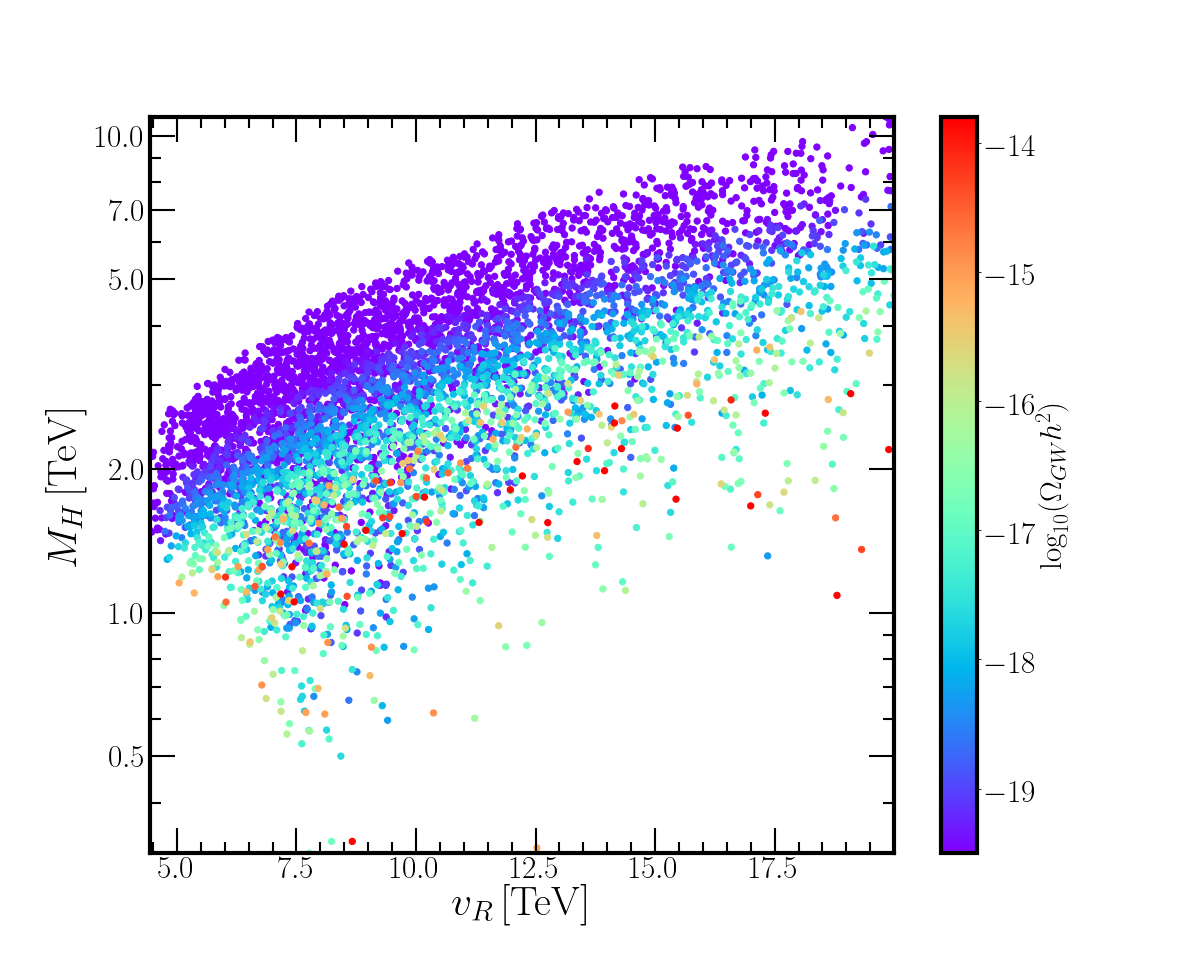}
  \end{minipage}
\caption{Left: Values of $\alpha$ versus $\frac{\beta}{H_\star}$ for the parameter points having detectable gravitational wave signal. Strong gravitational wave signal prefers large value of $\alpha$ and small value of $\frac{\beta}{H_\star}$. Right: Mass of heavy Higgs $M_H$ versus $v_R$ colour-coded with the maximum amplitude of $\Omega_{\mathrm{GW}}h^2$ for the points with detectable gravitational wave signal in future detectors. A strong gravitational wave signal prefers values of $M_H$ to be around $1$ TeV, but even slightly larger scalar masses can generate detectable gravitational waves. }
  \label{fig:abMHvR}
\end{figure}

Gravitational waves produced by the collisions of the bubble walls can be treated using the envelope approximation with their contribution to the total spectrum described by~\cite{Kosowsky:1991ua,Jinno:2016vai},
\begin{equation}
    \Omega_{\mathrm{coll}}h^2=\frac {1.67 \times 10^{-5}}{(\beta/H)^2}\left ( \frac {k_{\mathrm{c}} \alpha}{1+\alpha} \right )^{2}\left ( \frac {100}{g_{*}} \right )^{1/3}\left ( \frac {0.11v_{\mathrm{w}}^3}{0.42+v_{\mathrm{w}}^2} \right )\frac {3.8(f/f_{\mathrm{env}})^{2.8}}{1+2.8(f/f_{\mathrm{env}})^{3.8}},
\end{equation}
where $k_{\mathrm{c}}$ is the efficiency factor of bubble collision given by \cite{Kamionkowski:1993fg},
\begin{equation}
    k_{\mathrm{c}}=\frac {0.715\alpha+\frac {4}{27}\sqrt{\frac{3\alpha}{2}}}{1+0.715\alpha},
\end{equation}
and the peak frequency $f_{\mathrm{env}}$ by~\cite{Huber:2008hg},
\begin{equation}
    f_{\mathrm{env}}=16.5\times 10^{-6}\left ( \frac {0.62}{1.8-0.1v_{\mathrm{w}}+v_{\mathrm{w}}^2} \right )\left ( \frac {\beta}{H} \right )\left ( \frac {T_n}{100GeV} \right )\left ( \frac {g_{*}}{100} \right )^{1/6} \mathrm{Hz}.
\end{equation}

To provide some quantitative information on the detectability of the gravitational wave signal in the detector \cite{Caprini:2015zlo}, we use the signal-to-noise ratio defined as
\begin{equation}
    \mathrm{SNR} = \sqrt{\tau \int_{f_{\mathrm{min}}}^{f_{\mathrm{max}}}{\mathrm{d} f\left[ \frac {\Omega_{\mathrm{GW}}(f)h^2}{\Omega_{\mathrm{sens}}(f)h^2} \right]^2}},\label{SNR}
\end{equation}
where $\tau$ is the runtime of the particular experiment and one integrates the ratio of the gravitational wave signal strength $\Omega_{\mathrm{GW}}(f)h^2$ and the effective strain noise power spectral density $\Omega_{\mathrm{sens}}(f)h^2$ over the range of frequencies, in which the chosen experiment is sensitive.

\begin{figure}[th!]
  \centering
  \begin{minipage}[b]{0.489\textwidth}
    \includegraphics[width=\textwidth]{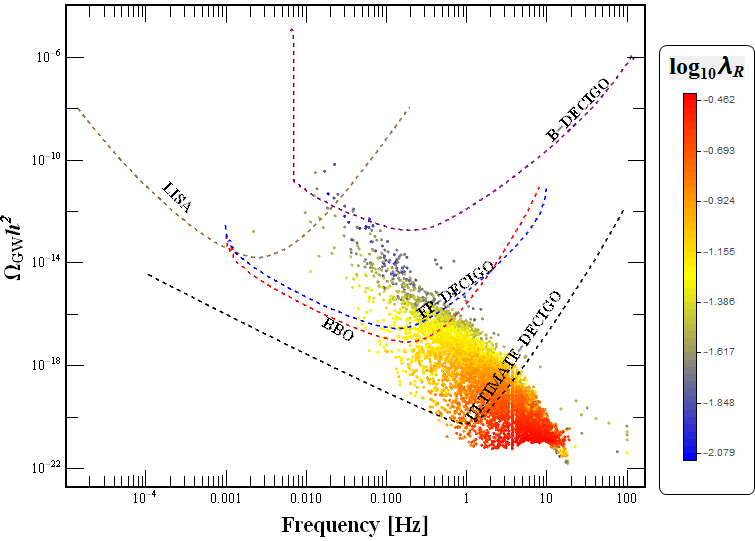}
  \end{minipage}
  \hfill
  \begin{minipage}[b]{0.495\textwidth}
    \includegraphics[width=\textwidth]{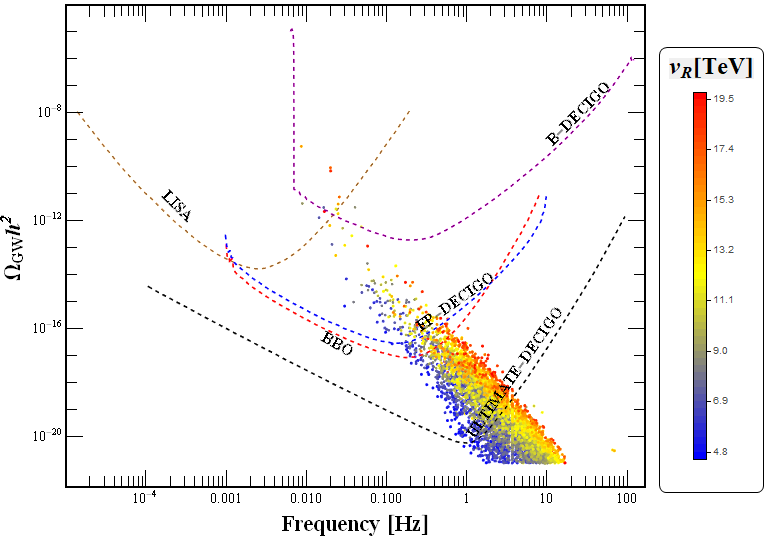}
  \end{minipage}
  \caption{Gravitational wave peak colour-coded with $\lambda_R$ (left) and $v_R$ (right). Generally, the smaller the value of  $\lambda_R$, the stronger signal is obtained. Thanks to the fact that the strongest signals are associated with lower peak frequencies, they can be detected by LISA. From the plot on the right, one can infer that for a given $\Omega_{\text{GW}}h^2$ smaller values of $v_R$ correspond to lower peak frequencies and vice versa.}
  \label{fig:GWscans}
\end{figure}

As mentioned in the previous section, our scan of the parameter space of the studied left-right symmetric model found that a strong first-order phase transition is preferably associated with small values of $\lambda_R$; therefore, in the subsequent analysis we focus on the subspace of scenarios satisfying this condition. When studying the correlation of parameters $\alpha$ and $\beta$ characterizing the phase transition, we find that, as expected, a strong gravitational signal is generally associated with large parameter $\alpha$, which in our case also corresponds to smaller values of $\beta$, see the left plot in Fig.~\ref{fig:abMHvR}. In the right plot of the same figure we show $M_H$ in connection to $v_R$ and the maximum amplitude of $\Omega_{\mathrm{GW}}(f)h^2$ suggesting that stronger signals prefer lower values of $M_H$; however, even larger scalar masses allow for production of detectable gravitational waves provided slightly larger $v_R$ is considered. The main results summarizing the numerical analysis of the gravitational wave signal in dependence on the free parameters of our model and in confrontation with the future experimental sensitivities are presented in Fig.~\ref{fig:GWscans}. Clearly, there is a non-negligible region of the parameter space giving gravitational wave signals, good part of which will be detectable by future space-based interferometers. In agreement with Fig.~\ref{fig:ptstrength}, strong signals typically correspond to small values of $\lambda_R$ and lower peak frequencies, which in turn allows for their detection by LISA. In general, the strength of the gravitational wave signal does not correlate with $v_R$, but smaller values of $v_R$ correspond to slightly lower peak frequencies and vice versa for a fixed value of $\Omega_{\mathrm{GW}}(f)h^2$.

\begin{figure}[t!]
  \centering
    \includegraphics[width=0.7\textwidth]{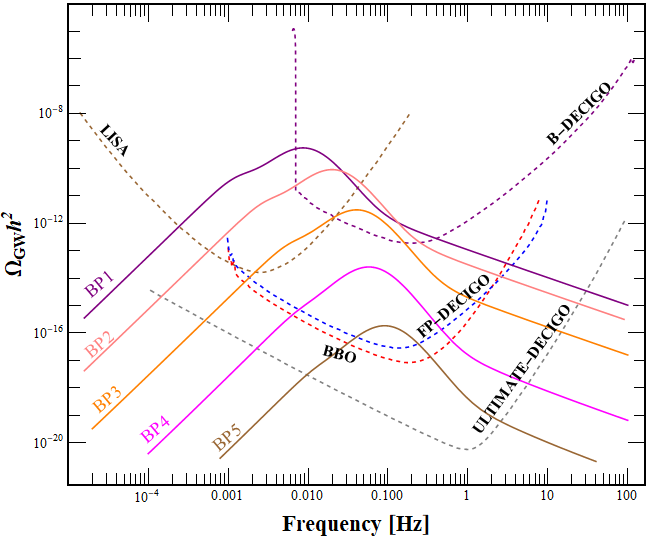}
  \caption{Stochastic gravitational wave as a function of frequency for the five selected benchmark points listed in Tab.~\ref{tab:1}.}
  \label{fig:gwbenchmark}
\end{figure}

Besides the generic scan of the parameter space, we present in~Fig.~\ref{fig:gwbenchmark} the resulting stochastic gravitational wave spectra in dependence on frequency for the five benchmark points listed in Tab.~\ref{tab:1} putting them in context with the expected sensitivities of several proposed space-based interferometers. As can be immediately inferred, all the five selected scenarios could be potentially detected by several future gravitational wave observatories with the most promising candidate being the first benchmark point BP1. While all the plotted spectra could be detected by FP-DECIGO, BBO and ULTIMATE-DECIGO experiments, only the first three benchmark points, BP1-BP3, would trigger a signal in B-DECIGO and LISA inteferometers with single-detector configurations. The corresponding SNRs are in all the cases well above the threshold value of $10$. Namely, taking $\tau=3$ years, the BP1, BP2 and BP3 can be probed in LISA with SNR larger than $10^6$, $10^5$ and $10^3$, respectively. The first stage of the DECIGO experiment, B-DECIGO, can probe BP1, BP2 and BP3 with all SNRs of order of $10^{4}-10^{5}$. Atlthough the SNRs of BP4 and BP5 are obviously smaller than 1, when LISA and B-DECIGO experiments are considered, these benchmark points can be safely probed by BBO and FP-DECIGO with SNRs in the range of $10^4-10^6$. The change of the slope on the right side of the spectra is caused by the fact that at larger frequencies the contribution from the bubble wall collisions becomes dominant over the one induced by sound waves, which decreases faster as $f \rightarrow \infty$. Similarly, the slight wiggle appearing to the left from the maximum of the gravitational wave spectra is due to the slight shift between peaks of gravitational wave contributions from sound waves and bubble wall collisions.

\begin{figure}[t!]
  \centering
    \includegraphics[width=1\textwidth]{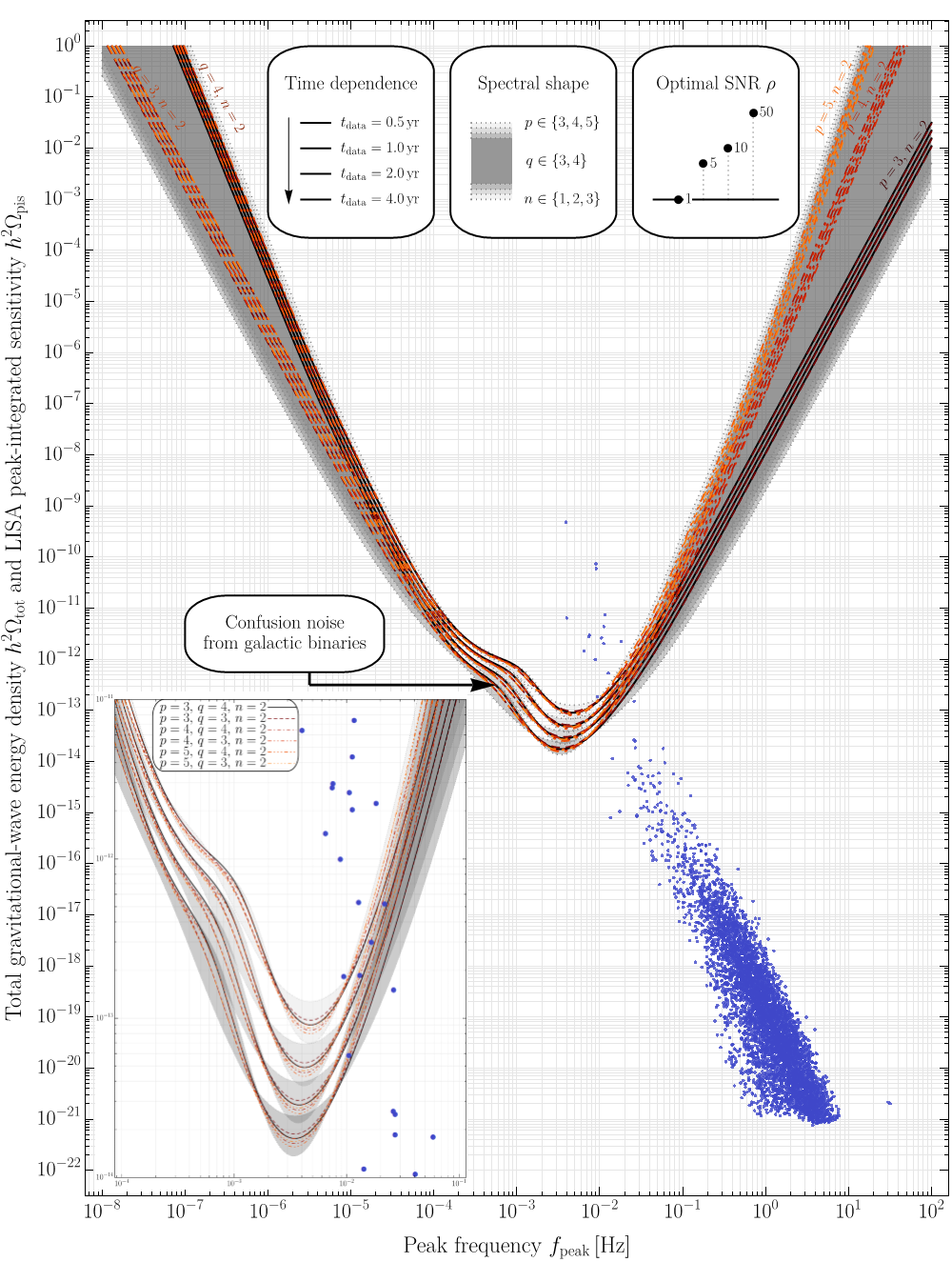}
  \caption{ LISA’s sensitivity to the acoustic GW signal from first order phase transition in our model. For details, see text.  }
  \label{fig:PISC}
\end{figure}

In plotting the GW signals in Fig.~\ref{fig:GWscans}, we have used the standard ansatz for the spectral shape function \cite{Caprini:2015zlo,Caprini:2019egz}. Recently, in Refs. \cite{Alanne:2019bsm,Schmitz:2020syl,Schmitz:2020rag} it is suggested that a broken power-law form for the spectral function is not always necessarily the best choice. In order to take into account for its
variability as well as uncertainty in the spectral function, a general class of broken power law is more suitable when compared with LISA sensitivities. A form of such a function is specified with a $(p, q, n)$ tuples; $p$ and $q$ here specify the range of power laws and $n$ identify the type of the peak in the spectrum (for details, see Refs. \cite{Alanne:2019bsm,Schmitz:2020syl,Schmitz:2020rag}).  The SNR we calculated in Eq.~\eqref{SNR} is the maximum theoretical possible value for the LISA's sensitivities. %They are obtained assuming an idealized auto-correlation measurement with perfect noise subtraction, assuming a stochastic, Gaussian, stationary, isotropic, and unpolarized GW background in the weak-signal regime \cite{Schmitz:2020rag, Schmitz:2020syl, Alanne:2019bsm}. 
A more realistic way of defining SNR is using LISA's peak integrated curves. Following the prescription described, for example in Ref. \cite{Schmitz:2020rag}, in Fig.~\ref{fig:PISC}, we draw the peak-integrated sensitivity plot for LISA’s sensitivity to acoustic GWs from cosmological phase transition of our model. 

The amplitude of the gravitational wave spectrum depends on the thermodynamic parameters of a first-order phase transition that are subject to theoretical uncertainties. Theoretical sources of uncertainties arise from renormalization scale dependence, gauge dependence, and the high temperature approximation used during computation. For discussions on the gauge dependence relevant for the calculation of
the spectrum of stochastic gravity waves, see for example Refs.  \cite{Dolan:1973qd, Patel:2011th, Wainwright:2011qy, Garny:2012cg}.  Estimation of these uncertainties is an active field of research, for recent studies in the SM effective field theory approach see for example Refs. \cite{Gould:2019qek,Croon:2020cgk,Gould:2021oba,Niemi:2021qvp,Schicho:2021gca}.

Here we briefly discuss the possibility of baryogenesis within this model. Due to the simplicity of the scalar sector, there is no CP-violating parameter present in this sector. However, baryogenesis may be generated utilizing CP-violating phases arising from the Yukawa sector. A detailed investigation is required to draw a conclusion, which is, however,  beyond the scope of this work. For related works along this line, see, e.g., \cite{Choi:1992wb, Mohapatra:1992pk,Frere:1992bd}.

%%%%%%%%%%%%%%%%%%%%%%%%%%%%%%%%%%%%%%%%%%%%%%%%%%%%%%%%%%%%
%%%%%%%%%%%%%%%%%%%%%%%%%%%%%%%%%%%%%%%%%%%%%%%%%%%%%%%%%%%%
\section{Collider Implications}
\label{col}
%%%%%%%%%%%%%%%%%%%%%%%%%%%%%%%%%%%%%%%%%%%%%%%%%%%%%%%%%%%%
%%%%%%%%%%%%%%%%%%%%%%%%%%%%%%%%%%%%%%%%%%%%%%%%%%%%%%%%%%%%
The emergence of new heavy gauge bosons ($W_R, Z_R$), scalar $H$, and vector-like fermions from the left-right symmetric universal seesaw framework would have several collider implications. This section therefore discusses some of the possible tests of our model. 

The heavy charged gauge boson $W_R$ inherits couplings with right-handed SM fermions, but the heavy neutral gauge boson $Z_R$ can communicate with both the left-handed and the right-handed SM fermions.
\begin{figure}[t!]
  \centering
  \begin{minipage}[b]{0.49\textwidth}
    \includegraphics[width=\textwidth]{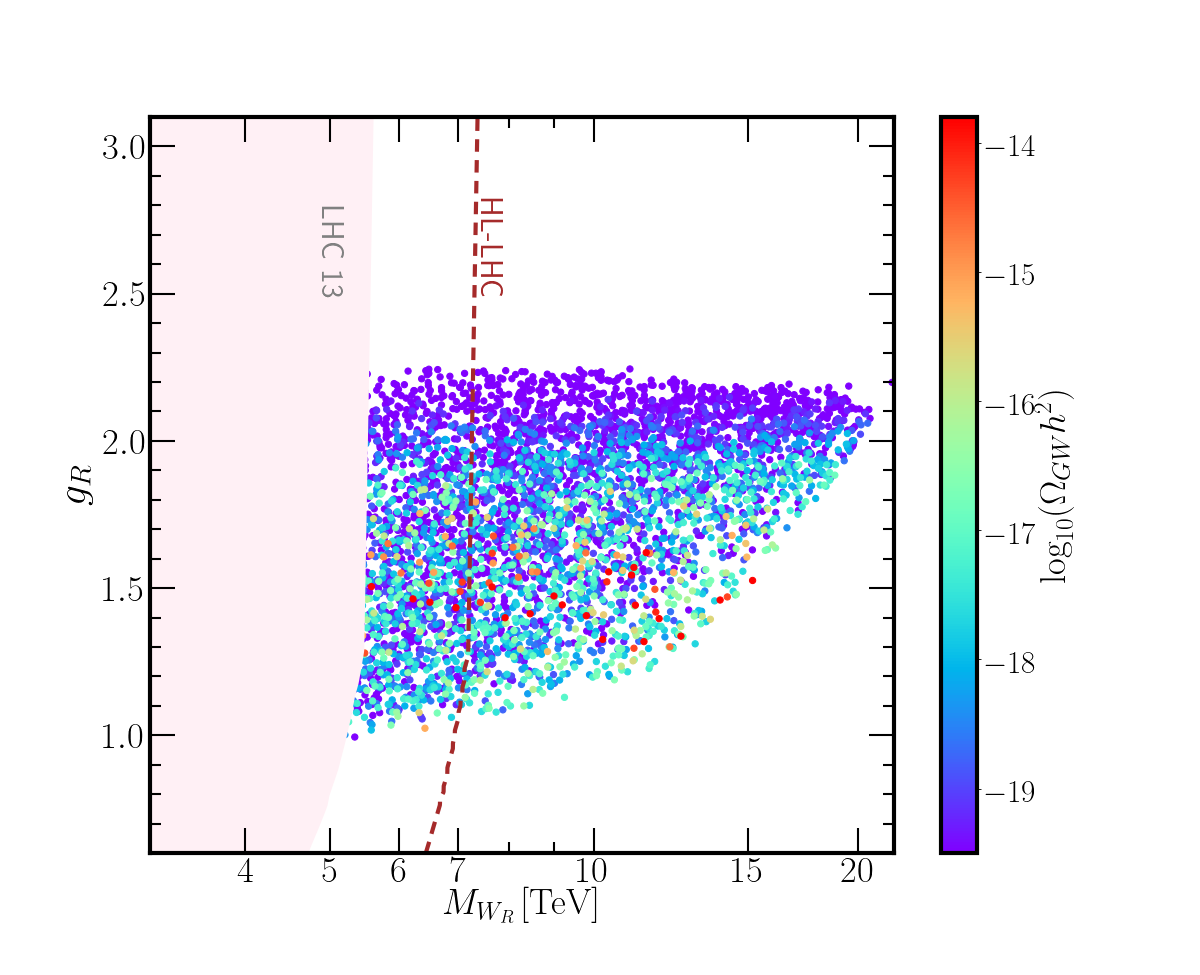}
  \end{minipage}
  \hfill
  \begin{minipage}[b]{0.49\textwidth}
    \includegraphics[width=\textwidth]{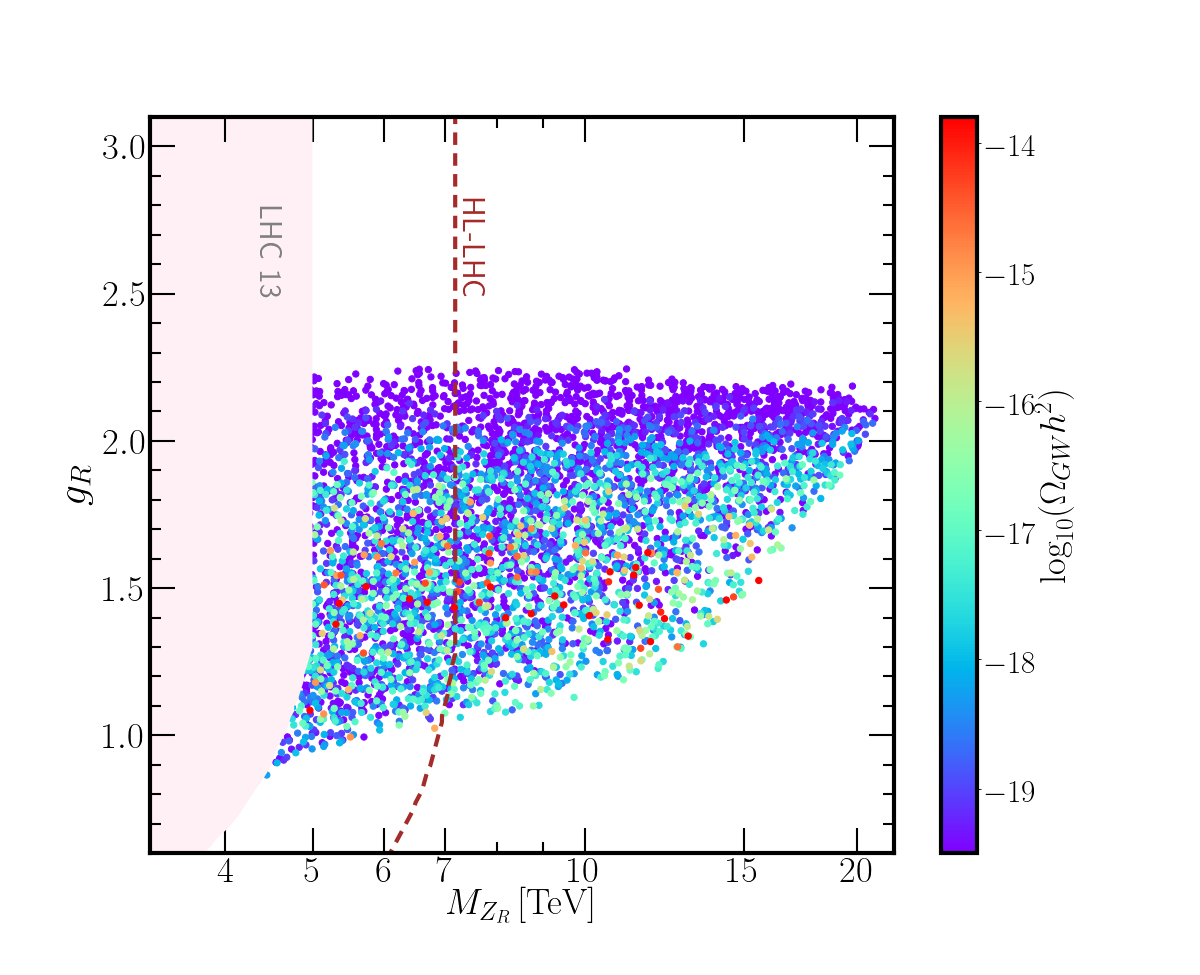}
  \end{minipage}
  \caption{Current collider limits and sensitivities for future collider experiments in the gauge coupling ($g_R$) versus the heavy gauge boson mass ($M_{W_R}$) plane. The light red shaded region indicates the current LHC limit, whereas the red dashed line indicates the future sensitivity for HL-LHC with a center of mass-energy $\sqrt{s} =14 $ TeV and an integrated luminosity of $\mathcal{L}=$ 3 ab$^{-1}$. Colour-coded points depict the maximum amplitude of $\Omega_{\mathrm{GW}}h^2$ for the scenarios associated with gravitational wave signal detectable by planned space-based interferometers. Note that these collider limits and sensitivities are based on particular assumptions, and they are not generic for this model. See text for details.}
  \label{collider}
\end{figure}
In the left-right symmetric model with triplet scalars, assuming $V_L=V_R$ (here, $V_L$ $(V_R)$ is the left-handed (right-handed) CKM mixing matrix), the limit on the mass of the $W_R$ vector boson, $M_{W_R}\geq 2.5~g_R$ TeV, must be satisfied to suppress the flavour violating interactions mediated by $W_R$. The most dangerous aspect is that such processes involve meson-antimeson oscillations, for example, $K^0_L-K^0_S$, $B^0_i-\overline B^0_i$ mixings ($i=d, s$) arising at one-loop level. As discussed earlier, a strong first-order phase transition in the model under consideration requires $g_R\sim 1.5$, which would lead to a bound of $M_{W_R}\geq 4$ TeV. Experimental limit on $W_R$ mass depends on the specific values of the Yukawa couplings as well as the masses of vector-like fermions. The bound quoted above, however, can be significantly relaxed with some judicious choice of the relevant Yukawa couplings; see for example Refs.~\cite{Babu:2018vrl,Frank:2018ifw}. Typically, both light and heavy fermions couple to $W_R$, which is a consequence of the seesaw formula. However, the seesaw formula breaks down if a texture corresponding to $\mathrm{det}(M_F)=0$ is chosen. For example, if $(M_U)_{11}=0$ is set in the diagonal $M_U$ matrix, then only the heavy $U$-quark couples to $W_R$, but not the light $u$-quark. In this case, the seesaw formula is not applicable to the first-generation up quark, contrary to the second and third generations. The same texture in the $M_D$ matrix will forbid light $d$-quark to couple to $W_R$, hence, significantly reducing collider bounds, for details see Ref.~\cite{Babu:2018vrl}. Couplings of fermions to $Z_R$ are also model specific and various different scenarios may emerge depending on the chosen flavour structure. For example, see Ref.~\cite{Babu:2018vrl}, where a bound of $M_{Z_R, W_R} \gtrsim 1$ TeV is obtained for a specific flavour structure. In this work, since the flavour structure is not determined from phenomenology, we adopt a conservative approach and take bounds on both $W_R$ and $Z_R$ masses similar to that of the conventional left-right symmetry model. Under this assumption, the most promising signal from heavy charged gauge boson $W_R$ will be $pp \to W_R \to l^{\pm} N \to l^{\pm} l^{\pm} jj $. There are dedicated searches for this lepton number violating signal~\cite{ATLAS:2018dcj}, which impose a strong bound on the ($g_R$ vs $M_{W_R}$) plane as shown by the light red colour shaded region in left panel of Fig.~\ref{collider}. Note that we assume the heavy right-handed neutrino mass to be 1 TeV for simplicity. It is important to mention that the resonant production rate of the heavy gauge bosons depends on the coupling $g_R$; however, the branching ratio to the above-mentioned mode does not have $g_R$ dependency, as we consider a negligible mixing between $W_L$ and $W_R$. The Fig.~\ref{collider} shows that $W_R$ mass up to 4.7 TeV is excluded from current collider data~\cite{ATLAS:2018dcj}, when $g_R = g_L$. We recast the current LHC limit to project the sensitivity for future HL-LHC searches with the center of mass energy $\sqrt{s} = 14$ TeV and an integrated luminosity of 3 ab$^{-1}$. It is quite interesting to see that a part of the parameter space for detectable gravitational wave signal can be complementarily probed by the HL-LHC experiment. For completeness, we present the collider complementarity for the $Z_R$ scenario as well in the right panel of Fig.~\ref{collider}. The most stringent bound comes from di-lepton resonant searches~\cite{ATLAS:2019erb} at the LHC and $Z_R$ mass is excluded up to $M_{Z_R} \gtrsim 5$ TeV. The searches for high mass phenomena in di-jet final states will also impose a bound on the model parameter space, but it is somewhat weaker than the di-lepton searches due to the large QCD background. Note that, throughout our analysis, we consider $Z_L - Z_R$ mixing to be small, and hence, the decay modes of $Z_R$ to $W^+ W^-$ or $Zh$ will be suppressed. As can be inferred from the right plot in Fig.~\ref{collider}, heavy $Z_R$ with mass up to $\sim$ 7 TeV can be probed by looking at dilepton resonant signal at the HL-LHC, which would also lead to exclusion of part of the parameter space giving detectable gravitational wave signals.

Throughout our study, we consider vector-like fermions to be very heavy. However, it is worth mentioning that the model parameter space allows these vector-like fermions to lie in the TeV or sub-TeV range. If that is the case, then it can be copiously pair produced at the collider via the Drell-Yan process and will subsequently lead to the smoking gun signatures of multi-leptons in association with jets or missing energy, which can be probed at the future collider experiments~\cite{Bhattiprolu:2019vdu,Chala:2018qdf}. For our analysis, we consider negligible mixing between the heavy Higgs and the SM Higgs as well. However, when turning the mixing on, the heavy Higgs can be produced at colliders giving the di-Higgs signal, which has been extensively investigated in the literature~\cite{Goncalves:2021egx,Alves:2019igs,Barman:2020ulr,Alves:2020bpi}. Again, we have to emphasize that the collider limits and future sensitivity are based on particular assumptions, which can be relaxed by choosing different benchmark points of the model.

%%%%%%%%%%%%%%%%%%%%%%%%%%%%%%%%%%%%%%%%%%%%%%%%%%%%%%%%%%%%%%%%%%%%
%%%%%%%%%%%%%%%%%%%%%%%%%%%%%%%%%%%%%%%%%%%%%%%%%%%%%%%%%%%%%%%%%%%%
\section{Summary and Conclusions}
\label{sec:Con}
%%%%%%%%%%%%%%%%%%%%%%%%%%%%%%%%%%%%%%%%%%%%%%%%%%%%%%%%%%%%%%%%%%%%
%%%%%%%%%%%%%%%%%%%%%%%%%%%%%%%%%%%%%%%%%%%%%%%%%%%%%%%%%%%%%%%%%%%%
In view of the increasingly stringent constraints on TeV and sub-TeV scale new physics imposed by current collider searches and other laboratory experiments, it becomes more and more important to explore new ways of probing physics at higher energy scales, and astrophysics and cosmology seem to be promising fields in this regard. One such possibility that emerged with the groundbreaking observations reported by LIGO and Virgo collaborations is to search for gravitational wave signals associated with new particle physics. A very intriguing source of gravitational waves linked to models beyond the SM could be first-order phase transitions connected to spontaneous symmetry breaking within particle physics models with extended gauge symmetries. Although these signals probably would not be detectable by the ground-based gravitational wave observatories, they could be within reach of future space missions, such as LISA, the launch of which is planned for 2037.

In this work, we have focused on cosmic phase transition associated with the left-right symmetry breaking in the model with the simplest Higgs sector and the universal seesaw mechanism generating the masses of the SM fermions. Performing a detailed numerical analysis, we have identified the region of the parameter space of the model allowing for first-order phase transition, and computed the associated stochastic gravitational wave signal. Studying the dependence of the characteristics of the gravitational wave production on the parameters of the model shows that particularly strong signals can be obtained for small values of the quartic right-handed doublet self-coupling $\lambda_R$, see Fig.~\ref{fig:ptstrength} or the left plot in Fig.~\ref{fig:GWscans}. At the same time, they require the value of the gauge coupling $g_R$ to be around $1$ or higher; therefore, the manifestly left-right symmetric scenario is disfavoured in our analysis. The heavy Higgs mass $M_H$ then generally acquires values of the order of a TeV in the benchmark points with detectable gravitational wave signals, which is supported by the right plot in Fig.~\ref{fig:abMHvR}. When confronting the gravitational wave peaks with the expected experimental sensitivities, we found that a good part of the studied parameter space would give signals detectable by planned space-based interferometers including LISA and B-DECIGO, as captured by Fig.~\ref{fig:GWscans}, where the colour-coding also depicts the dependence of the parameter $\lambda_R$ and the left-right symmetry breaking scale $v_R$. To demonstrate explicitly the shape of the stochastic gravitational wave spectra obtained in our analysis we selected five interesting benchmark points characterized by values of parameter listed in Tab.~\ref{tab:1} and plotted the corresponding curves in Fig.~\ref{fig:gwbenchmark}. The overall shape of the spectra is given by the contributions from sound waves and bubble wall collisions, with the former one being dominant. The SNRs calculated for the five benchmark points confirm their detectability by BBO, FP-DECIGO, and ULTIMATE-DECIGO experiments with the two strongest signals safely reaching also sensitivities of B-DECIGO and LISA. Furthermore, LISA’s expected sensitivities to these gravitational wave signals based on the concept of peak-integrated sensitivity curves are presented in Fig.~\ref{fig:PISC}, which clearly shows the potential of this model to be observable in the LISA mission.

In addition to the detailed analysis of the gravitational wave signals induced by the strong first-order phase transition, we have also examined the prospect of exploring the described model at existing and future colliders discussing the potential complementarity. It is, however, important to note that the upcoming gravitational wave experiments would also allow investigating the parameter space in our model that remains beyond the reach of collider searches.

%%%%%%%%%%%%%%%%%%%%%%%%%%%%%%%%%%%%%%%%%%%%%%%%%%%%%%%%%%%%%%%%%%%%
%%%%%%%%%%%%%%%%%%%%%%%%%%%%%%%%%%%%%%%%%%%%%%%%%%%%%%%%%%%%%%%%%%%%
\section*{Acknowledgments}
%%%%%%%%%%%%%%%%%%%%%%%%%%%%%%%%%%%%%%%%%%%%%%%%%%%%%%%%%%%%%%%%%%%%
%%%%%%%%%%%%%%%%%%%%%%%%%%%%%%%%%%%%%%%%%%%%%%%%%%%%%%%%%%%%%%%%%%%%
Authors thank K.S. Babu, R. Dcruz and G. Chauhan for useful discussions. We thank Kai Schmitz for email correspondence and for providing us with a Mathematica notebook, which we have used to make Fig.~\ref{fig:PISC}.  Part of the computing for this project was performed at the High Performance Computing Center at Oklahoma State University, supported in part through the National Science Foundation grant OAC-1531128.  L.G. acknowledges support from the National Science Foundation, Grant PHY-1630782, and to the Heising-Simons Foundation, Grant 2017-228. The work of A.K.\ was in part supported by US Department of Energy Grant Number DE-SC 0016013. The work of S.S.\ has been supported by the Swiss National Science Foundation. 

%%%%%%%%%%%%%%%%%%%%%%%%%%%%%%%%%%%%%%%%%%%%%
%%%%%%%%%%%%%%%%%%%%%%%%%%%%%%%%%%%%%%%%%%%%%
\bibliographystyle{utphys}
{\footnotesize
\bibliography{reference}}
\end{document}